\documentclass[conference]{IEEEtran}
\IEEEoverridecommandlockouts

\usepackage{cite}
\usepackage{amsmath,amssymb,amsfonts}
\usepackage{algorithmic}
\usepackage{graphicx}
\usepackage{textcomp}
\usepackage{xcolor}

\usepackage{comment}
\usepackage{array}
\usepackage{booktabs}
\usepackage{multirow}
\usepackage{subcaption}

\def\BibTeX{{\rm B\kern-.05em{\sc i\kern-.025em b}\kern-.08em
    T\kern-.1667em\lower.7ex\hbox{E}\kern-.125emX}}
\begin{document}

\title{Macro-Level Correlational Analysis of Mental Disorders: Economy, Education, Society, and Technology Development}

\author{\IEEEauthorblockN{Yingzhi Tao\IEEEauthorrefmark{1}\thanks{*All authors contributed equally.}}
 	\IEEEauthorblockA{\textit{Department of Informatics} \\ 
 		\textit{King's College London} \\ 
 		London, United Kingdom \\ 
 		yingzhi.tao@kcl.ac.uk\\
 		ORCID: 0009-0006-2538-6052}
	\and
 	\IEEEauthorblockN{Chang Yang}
 	\IEEEauthorblockA{\textit{Department of Informatics} \\ 
 		\textit{King's College London} \\ 
 		London, United Kingdom \\ 
 		chang.2.yang@kcl.ac.uk\\
 		ORCID: 0009-0000-8984-9796}
}

\maketitle

\begin{abstract}
This paper quantifies the age-stratified global burden of four mental disorders in 27 regions from 1990 to 2021 using GBD 2021. To put it in detail, it links the age-standardized years of disability adjustment with 18 world development indicators across economic, educational, social and information technology sectors. Then, by means of Pearson correlation, mutual information, Granger causality and maximum information coefficient and other methods, the linear, nonlinear and lagged dependency relationships were evaluated. After research, it was found that there is a very prominent spatio-temporal heterogeneity among young people aged 20 to 39, and the coupling relationship is stronger. From the overall situation, education corresponds to a low burden. Unemployment corresponds to a high burden. Through lag analysis, it can be known that the influence time of economic and technological factors is relatively short, while that of educational factors is relatively long. These results highlight the macro determinants that play a role at different time scales and also provide population-level references for verifying computational mental health models and for intervention measures in specific regions and for specific ages.
\end{abstract}


\section{Introduction}

Mental health disorders are a particularly significant global health challenge in the 21st century. This challenge affects over one billion people worldwide and has largely led to an increase in morbidity and mortality rates globally \cite{mental_health_global_burden}. Recent reports show that mental health disorders have become a major factor causing disability worldwide, and it is necessary to take prompt action. And comprehensively understand its determining factors \cite{mental_health_inequality}. In all regions around the world, the disease burden caused by mental health disorders is increasing rapidly. The development trend of early-onset mental health conditions \cite{early_onset_mental_health} is very worrying. Understanding the distribution pattern of mental health burden in time and space, as well as the differences between early-onset and late-onset diseases, is crucial for formulating targeted prevention strategies and public health intervention measures.

The causes of mental health disorders are multifaceted, involving genetic susceptibility, environmental factors, and the influence of socio-economic factors. Among these factors, socio-economic determinants have received increasing attention due to their potential role in regulating mental health outcomes \cite{social_determinants_mental_health}, such as economic development, educational attainment, social stability, and digital connectivity. All these aspects play a crucial role in mental health outcomes \cite{economic_mental_health}. The education system will have an impact on mental health literacy and people's access to care. Economic stability will affect the stress level of mental health services \cite{education_mental_health} and the supply of resources. The development of information technology will have an impact on people's access to mental health resources and digital mental health intervention measures \cite{digital_mental_health}.

This paper utilized comprehensive data from the "Global Burden of Disease Study 2021" to conduct a systematic assessment of the global burden of mental health disorders among different age groups. We particularly focused on two distinct age groups: one is young people, aged between 20 and 39, and the other is elderly people, that is, those over 40 years old. To investigate the differences in burden patterns and socio-economic determinants between early-onset and late-onset mental health conditions. Recently, some studies \cite{early_onset_mental_health} have highlighted that early-onset mental health disorders can have a significant impact on long-term outcomes, and have also proposed the necessity of formulating intervention strategies tailored to specific age groups.

The rationale for this age-based stratification stems from the growing recognition that early-onset and later-onset mental health conditions may have different etiological pathways, risk factor profiles, and clinical outcomes \cite{mental_health_global_burden}. By analyzing comprehensive burden metrics including prevalence, incidence, mortality, and DALYs for four major mental health disorders (bipolar disorder, anxiety disorders, schizophrenia, and depressive disorders), along with socioeconomic determinants from the World Bank database, we aim to provide evidence-based insights for developing age-specific prevention and management strategies. This approach is particularly relevant given the increasing global burden of mental health disorders and the role of socioeconomic factors in mental health outcomes \cite{mental_health_inequality}.

\section{Dataset}

\subsection{Data Sources}

Our analysis draws from two primary data sources, each providing comprehensive global coverage and standardized methodologies for data collection and reporting.

\subsubsection{Mental Health Data}

The data on mental health outcomes is derived from the Global Burden of Disease Study 2021 (GBD 2021), which provided a comprehensive analysis of 204 countries and regions \cite{gbd2021}. From this study, we extracted DALYs data for four major mental health disorders stratified by age group. The age groups here are divided into two groups: 20-39 years old and over 40 years old. The four main mental health disorders are bipolar disorder, anxiety disorder, schizophrenia and depression. Moreover, all the extracted data were processed for age standardization using the GBD world standard population.

DALYs (Disability-Adjusted Life Years) represent the sum of years lost due to premature death (YLLs) and years lived with disability (YLDs), defined as years of healthy life lost. The mathematical formulation is:

\begin{equation}
	\text{DALYs} = \text{YLLs} + \text{YLDs}
\end{equation}

\textbf{YLLs (Years of Life Lost)} are calculated as the multiplication of deaths and a standard life expectancy at the age of death:

\begin{equation}
	\text{YLLs} = \sum_{a} D_a \times L_a
\end{equation}

where $D_a$ represents the number of deaths at age $a$, and $L_a$ is the standard life expectancy at age $a$. The standard life expectancy is derived from a life table containing the lowest observed mortality rate at each age in populations greater than 5 million.

\textbf{YLDs (Years Lived with Disability)} represent years lived with any short-term or long-term health loss weighted for severity by disability weights:

\begin{equation}
	\text{YLDs} = \sum_{a} P_a \times DW_a
\end{equation}

where $P_a$ represents the prevalence of the condition at age $a$, and $DW_a$ is the disability weight for the condition at age $a$. Disability weights range from 0 (perfect health) to 1 (death), reflecting the severity of health loss.

\subsubsection{Socioeconomic and Development Data}

The socio-economic indicators selected in this paper are obtained from the World Bank's World Development Indicators (WDI) database \cite{worldbank2023}. These indicators cover four major aspects: economic development, educational attainment, social stability, and the adoption of information technology, involving a total of 19 variables. These indicators were selected based on their theoretical relevance to mental health outcomes and temporal data availability for correlation analysis.

\subsection{Variable Definitions and Selection Criteria}

\subsubsection{Dependent Variables}

This paper analyzes four major mental health disorders suffered by two age groups, namely bipolar disorder, anxiety disorder, schizophrenia and depression. These two age groups are young people, that is, people aged 20 to 39, and the elderly, that is, people over 40 \cite{early_onset_mental_health}. Such stratification is consistent with the early-onset mental health conditions that occur before the age of 40. The main outcome measure is disability-adjusted life years, which comprehensively assesses the burden of disease by combining the impacts of mortality and morbidity. All these indicators have been age-standardized to ensure comparability across different countries and at different times.

The time span of this dataset reaches 32 years (1990-2021). Such a long time span provides sufficient time range coverage for Granger causality analysis and trend assessment. All variables are age-standardized, and these variables are provided every other year. In this way, Then a comprehensive temporal analysis of the relationship between socio-economic and mental health can be conducted.

This paper incorporates a total of 19 socio-economic variables in four categories, which belong to different categories. The first category is economic indicators, including GDP, per capita GDP, inflation, and employment situations in agriculture, industry, and services. The second category is educational indicators, involving enrollment rates in primary schools, middle schools, and universities, as well as educational expenditures. The third category is social indicators, including life expectancy, unemployment rate and the incidence of malnutrition. The fourth category is information technology indicators, including Internet usage, mobile users, broadband, exports of information and communication technology products and secure servers. The selection of these variables is because they have a certain relationship with the mental health outcomes throughout the study period, and also takes into account the availability of the data (1990-2021).

\subsection{Study Variables}

Tables \ref{tab:dependent_variables} and \ref{tab:independent_variables} present the variables analyzed using our four-method framework, with DALYs as the sole outcome measure for comprehensive correlation analysis.

\begin{table}[htbp]
	\caption{Dependent Variables (Mental Health Outcomes)}
	\begin{center}
		\begin{tabular}{p{1.5cm}p{3.2cm}p{1.8cm}}
			\toprule
			\textbf{Category} & \textbf{Variable Name} & \textbf{Age Group} \\
			\midrule
			\multirow{8}{*}{\parbox{1.5cm}{\centering Mental Health \\ Disorders}} & Bipolar Disorder & 20-39 years \\
			& Bipolar Disorder & 40+ years \\
			& Anxiety Disorders & 20-39 years \\
			& Anxiety Disorders & 40+ years \\
			& Schizophrenia & 20-39 years \\
			& Schizophrenia & 40+ years \\
			& Depressive Disorders & 20-39 years \\
			& Depressive Disorders & 40+ years \\
			\bottomrule
		\end{tabular}
		\label{tab:dependent_variables}
	\end{center}
\end{table}

This table contains the mapping from full variable names to short codes for all factors. The codes are used in subsequent analysis and visualization to maintain clarity and readability.

\newcolumntype{C}[1]{>{\centering\arraybackslash}p{#1}} 

\begin{table}[htbp]
	\caption{Independent Variables (Socioeconomic Indicators)}
	\label{tab:independent_variables}
	\begin{center}
		\begin{tabular}{C{1.0cm}C{0.4cm}C{3.5cm}C{2.4cm}}
			\toprule
			\textbf{Category} & \textbf{Code} & \textbf{Variable Name} & \textbf{Units} \\
			\midrule
			\multirow{6}{*}{\parbox{\linewidth}{\centering Economic}} & E1 & GDP & Current US\$ \\
			& E2 & GDP per capita & Current US\$ \\
			& E3 & Inflation, consumer prices & \% \\
			& E4 & Employment in industry & \% \\
			& E5 & Employment in services & \% \\
			& E6 & Employment in agriculture & \% \\
			\cmidrule{1-4}
			\multirow{4}{*}{\parbox{\linewidth}{\centering Education}} & ED1 & School enrollment, primary & \% \\
			& ED2 & School enrollment, secondary & \% \\
			& ED3 & School enrollment, tertiary & \% \\
			& ED4 & Government expenditure on education, total & \% \\
			\cmidrule{1-4}
			\multirow{3}{*}{\parbox{\linewidth}{\centering Society}} & S1 & Life expectancy at birth, total & Years \\
			& S2 & Unemployment, total & \% \\
			& S3 & Prevalence of undernourishment & \% \\
			\cmidrule{1-4}
			\multirow{5}{*}{\parbox{\linewidth}{\centering Technology}} & T1 & Individuals using the Internet & \% \\
			& T2 & Mobile cellular subscriptions & per 100 people \\
			& T3 & Fixed broadband subscriptions & per 100 people \\
			& T4 & Secure Internet servers & per 1 million people \\
			& T5 & ICT goods exports & \% \\
			\bottomrule
		\end{tabular}
	\end{center}
\end{table}

\begin{table*}[htbp]
	\caption{Comprehensive Dataset: Socioeconomic Indicators by Year (1991-2023)}
	\begin{center}
		\footnotesize
		\setlength{\tabcolsep}{8pt}
		\begin{tabular}{c|ccc|cccc|ccc|ccccc}
			\toprule
			\multirow{2}{*}{\textbf{Year}} & \multicolumn{3}{c|}{\textbf{Economic}} & \multicolumn{4}{c|}{\textbf{Education}} & \multicolumn{3}{c|}{\textbf{Social}} & \multicolumn{5}{c}{\textbf{Technology}} \\
			\cmidrule{2-16}
			& E1 & E2 & E3 & ED1 & ED2 & ED3 & ED4 & S1 & S2 & S3 & T1 & T2 & T3 & T4 & T5 \\
			\midrule
			1991 & 23.9 & 4.4 & 9.0 & 97.9 & 51.5 & 13.7 & - & 65.3 & 5.1 & - & - & 0.3 & - & - & - \\
			1992 & 25.5 & 4.7 & 7.6 & 97.3 & 52.7 & 14.0 & - & 65.5 & 5.3 & - & - & 0.4 & - & - & - \\
			1993 & 26.0 & 4.7 & 7.1 & 97.4 & 54.0 & 14.4 & - & 65.7 & 5.6 & - & - & 0.6 & - & - & - \\
			1994 & 27.9 & 4.9 & 10.2 & 97.6 & 54.7 & 15.0 & - & 66.0 & 5.8 & - & - & 1.0 & - & - & - \\
			1995 & 31.2 & 5.4 & 9.1 & 97.6 & 55.3 & 15.6 & - & 66.2 & 5.9 & - & - & 1.6 & - & - & - \\
			1996 & 31.9 & 5.5 & 6.5 & 97.2 & 56.1 & 16.3 & - & 66.5 & 6.0 & - & - & 2.5 & - & - & - \\
			1997 & 31.8 & 5.4 & 5.6 & 97.5 & 56.9 & 17.0 & - & 66.8 & 6.0 & - & - & 3.7 & - & - & - \\
			1998 & 31.7 & 5.3 & 5.1 & 97.7 & 57.3 & 17.5 & - & 67.1 & 6.2 & - & - & 5.3 & - & - & - \\
			1999 & 32.8 & 5.4 & 3.0 & 98.4 & 57.7 & 18.4 & 4.1 & 67.3 & 6.3 & - & - & 8.1 & - & - & - \\
			2000 & 33.9 & 5.5 & 3.4 & 98.9 & 58.5 & 19.0 & 3.9 & 67.6 & 6.1 & - & - & 12.1 & - & - & 15.1 \\
			2001 & 33.7 & 5.4 & 3.8 & 100.3 & 60.3 & 20.2 & 4.0 & 67.9 & 6.2 & 13.0 & - & 15.4 & - & - & 14.6 \\
			2002 & 35.0 & 5.5 & 2.9 & 100.3 & 62.0 & 21.5 & 4.0 & 68.2 & 6.4 & 13.0 & - & 18.8 & - & - & 14.8 \\
			2003 & 39.2 & 6.1 & 3.0 & 101.9 & 62.8 & 22.6 & 4.2 & 68.5 & 6.5 & 12.9 & - & 22.3 & - & - & 14.9 \\
			2004 & 44.2 & 6.8 & 3.5 & 102.6 & 63.8 & 23.6 & 4.1 & 68.8 & 6.3 & 12.6 & - & 26.0 & - & - & 15.2 \\
			2005 & 47.8 & 7.3 & 4.1 & 102.9 & 64.8 & 24.3 & 4.1 & 69.1 & 6.2 & 12.0 & 15.6 & 33.9 & 3.4 & - & 14.3 \\
			2006 & 51.9 & 7.8 & 4.3 & 103.3 & 65.5 & 25.4 & 4.2 & 69.5 & 6.0 & 11.2 & 17.2 & 41.7 & 4.3 & - & 14.2 \\
			2007 & 58.4 & 8.7 & 4.8 & 104.0 & 67.3 & 26.4 & 4.1 & 69.8 & 5.8 & 10.3 & 20.2 & 50.6 & 5.2 & - & 13.1 \\
			2008 & 64.2 & 9.4 & 8.9 & 104.2 & 68.8 & 27.4 & 4.3 & 70.0 & 5.8 & 9.6 & 22.8 & 59.7 & 6.1 & - & 12.2 \\
			2009 & 60.9 & 8.8 & 2.9 & 103.9 & 70.0 & 28.4 & 4.5 & 70.4 & 6.4 & 9.1 & 25.3 & 68.0 & 6.9 & - & 13.1 \\
			2010 & 66.7 & 9.5 & 3.3 & 103.7 & 71.6 & 29.7 & 4.1 & 70.7 & 6.3 & 8.7 & 28.4 & 76.6 & 7.6 & 185.2 & 12.9 \\
			2011 & 74.2 & 10.5 & 4.8 & 103.7 & 72.9 & 31.4 & 4.1 & 71.0 & 6.2 & 8.3 & 30.9 & 84.2 & 8.6 & 236.1 & 11.6 \\
			2012 & 75.9 & 10.6 & 3.7 & 103.8 & 73.5 & 32.5 & 4.2 & 71.3 & 6.2 & 8.0 & 33.3 & 88.5 & 9.2 & 321.0 & 11.5 \\
			2013 & 78.1 & 10.7 & 2.7 & 103.9 & 74.9 & 33.4 & 4.3 & 71.5 & 6.1 & 7.7 & 35.3 & 93.1 & 9.7 & 365.9 & 11.3 \\
			2014 & 80.2 & 10.9 & 2.4 & 102.6 & 75.7 & 35.9 & 4.3 & 71.8 & 6.0 & 7.6 & 37.4 & 96.7 & 10.1 & 444.4 & 11.4 \\
			2015 & 75.7 & 10.2 & 1.4 & 102.0 & 75.7 & 37.0 & 4.2 & 72.0 & 6.0 & 7.5 & 39.8 & 96.1 & 11.3 & 565.5 & 11.9 \\
			2016 & 77.0 & 10.2 & 1.6 & 103.1 & 75.8 & 37.5 & 4.2 & 72.2 & 6.0 & 7.5 & 42.8 & 99.3 & 12.2 & 1249.8 & 12.1 \\
			2017 & 82.0 & 10.8 & 2.3 & 102.7 & 75.5 & 37.9 & 4.2 & 72.4 & 5.9 & 7.3 & 45.2 & 101.4 & 13.5 & 3470.1 & 12.3 \\
			2018 & 87.2 & 11.3 & 2.4 & 100.1 & 76.0 & 38.3 & 4.1 & 72.6 & 5.8 & 7.3 & 48.5 & 103.1 & 14.0 & 6087.1 & 12.5 \\
			2019 & 88.5 & 11.4 & 2.2 & 100.2 & 76.3 & 39.1 & 4.1 & 72.9 & 5.6 & 7.8 & 52.9 & 105.9 & 14.7 & 9854.2 & 12.5 \\
			2020 & 86.1 & 11.0 & 1.9 & 100.7 & 76.8 & 40.1 & 4.4 & 72.2 & 6.6 & 8.3 & 58.6 & 104.9 & 15.6 & 11366.7 & 13.4 \\
			2021 & 98.4 & 12.4 & 3.5 & 100.5 & 77.5 & 41.4 & 4.2 & 71.2 & 6.1 & 8.9 & 61.7 & 106.7 & 16.8 & 12808.2 & 13.2 \\
			2022 & 102.4 & 12.8 & 7.9 & 101.7 & 77.6 & 42.4 & 3.8 & 73.0 & 5.3 & 9.1 & 63.7 & 108.1 & 17.7 & 14527.7 & 11.9 \\
			2023 & 107.0 & 13.3 & 5.9 & 101.8 & 77.1 & 43.3 & - & 73.3 & 4.9 & - & 65.4 & 109.4 & 18.6 & 15466.2 & - \\
			\bottomrule
		\end{tabular}
		\label{tab:comprehensive_data}
	\end{center}
	\tiny
	\textbf{Note:} E1=GDP (trillion US\$), E2=GDP per capita (thousand US\$), E3=Inflation (\%); ED1-ED4=Education indicators (\%); S1=Life expectancy (years), S2=Unemployment (\%), S3=Undernourishment (\%); T1=Internet users (\%), T2=Mobile subscriptions (per 100), T3=Broadband (per 100), T4=Secure servers (per million), T5=ICT exports (\%). Missing values indicated by '-'.
\end{table*}

\subsection{Units and Measurements}

All variables use standardized international units for comparability. Mental health indicators are measured as dimensionless DALYs. Economic indicators include GDP and GDP per capita (current US\$), inflation rates (\%), and employment distribution (\%). Educational and social indicators are measured as percentages, while information technology indicators use percentages or per capita rates.

\section{Methods}
\subsection{Pearson's Correlation Coefficient}

The Pearson correlation coefficient is a parameter used to measure the linear relationship between two variables. It is defined as the product of the covariances of the two variables divided by their standard deviations. Its formula is as follows:

\begin{equation}
	r_{XY} = \frac{\sum_{i=1}^{n} (X_i - \overline{X})(Y_i - \overline{Y})}{\sqrt{\sum_{i=1}^{n} (X_i - \overline{X})^2 \sum_{i=1}^{n} (Y_i - \overline{Y})^2}} \label{eq:pearson}
\end{equation}

where \( r_{XY} \) represents the Pearson correlation coefficient between \( X \) and \( Y \), \( X_i \) and \( Y_i \) are the individual data points for variables \( X \) and \( Y \), and \( \overline{X} \) and \( \overline{Y} \) are the means of the variables. This coefficient takes values between -1 and 1, where 1 indicates a perfect positive linear correlation, -1 indicates a perfect negative linear correlation, and 0 indicates no linear correlation.

Pearson's method is relatively simple to operate and has a high computational efficiency. It provides a standardized measurement method for linear correlations. However, this method has its drawbacks. It is particularly sensitive to outliers and assumes binary normality, which makes it unable to detect nonlinear relationships or distinguish causal relationships \cite{pearson1896mathematical}.

\subsection{Mutual Information}

Mutual information is a non-parametric measure from information theory that quantifies the amount of information shared by two variables, capable of capturing both linear and non-linear relationships. The mutual information between two variables \( X \) and \( Y \) is defined as:

\begin{equation}
	I(X; Y) = \sum_{x \in X} \sum_{y \in Y} p(x, y) \log \frac{p(x, y)}{p(x)p(y)} \label{eq:mutual}
\end{equation}

where \( p(x, y) \) is the joint probability distribution of \( X \) and \( Y \), and \( p(x) \) and \( p(y) \) are the marginal probability distributions of \( X \) and \( Y \), respectively.

Mutual information is non-parametric and has relatively good robustness to data features. It can capture both linear and nonlinear relationships between data. However, mutual information is rather computationally demanding and requires discretization of continuous variables \cite{shannon1948mathematical}.

\subsection{Granger Causality Test}

The Granger Causality test is a statistical hypothesis test used to determine whether one time series can predict another, based on the premise that if variable \( X \) Granger-causes variable \( Y \), past values of \( X \) will contain information that helps predict \( Y \). The null hypothesis is that \( X \) does not Granger-cause \( Y \). The test statistic is calculated as:

\begin{equation}
	F = \frac{(RSS_{r} - RSS_{ur}) / k}{RSS_{ur} / (n - k - 1)} \label{eq:granger}
\end{equation}

where \( F \) is the test statistic, \( RSS_{r} \) is the residual sum of squares from the restricted model (without lag of \( X \)), \( RSS_{ur} \) is the residual sum of squares from the unrestricted model (with lag of \( X \)), \( k \) is the number of parameters, and \( n \) is the number of observations.

Granger causality can detect time-dependent and directional relationships, but it has a prerequisite: assuming the data is stationary and there is also a need for sufficient observation. However, it should be noted that the relationship detected by Granger causality does not necessarily represent a true causal relationship \cite{granger1969investigating}.

\subsection{Maximal Information Coefficient (MIC)}

The Maximal Information Coefficient (MIC) is a non-parametric method designed to measure both linear and non-linear dependencies between two variables, part of the Maximal Information-based Nonparametric Exploration (MINE) framework. The MIC is defined as the maximum value of the normalized mutual information, calculated by discretizing the variables into bins:

\begin{equation}
	\text{MIC}(X, Y) = \max_{b_1, b_2} I(X; Y) / \max \left( H(X), H(Y) \right) \label{eq:mic}
\end{equation}

where \( I(X; Y) \) is the mutual information between \( X \) and \( Y \), \( H(X) \) and \( H(Y) \) are the entropy of \( X \) and \( Y \), respectively, and \( b_1 \) and \( b_2 \) are the bins used to discretize the data.

MIC provides standardized values ranging from 0 to 1. It can compare different relationship types and detect both linear and nonlinear patterns. However, its computational complexity is relatively high and it may be sensitive to discretization bias \cite{reshef2011maximal}.

\section{Experiments and Analysis}

In this section, this paper employs four distinct correlation methods to conduct a comprehensive analysis of the relationship between socio-economic factors and mental health outcomes among different age groups. The multi-method approach we adopt enables us to identify both linear and nonlinear relationships, as well as understand time dependence. And the complex interaction between socio-economic determinants and mental health burdens.

\subsection{Pearson Correlation Analysis}

Pearson correlation analysis can provide us with some insights into the linear relationship between socio-economic factors and mental health outcomes. This parametric approach can effectively capture the direct linear association between them, making it particularly suitable for determining the direct relationship between economic development and mental health burden in different age groups.

\begin{figure*}[htbp]
	\centering
\includegraphics[width=\textwidth]{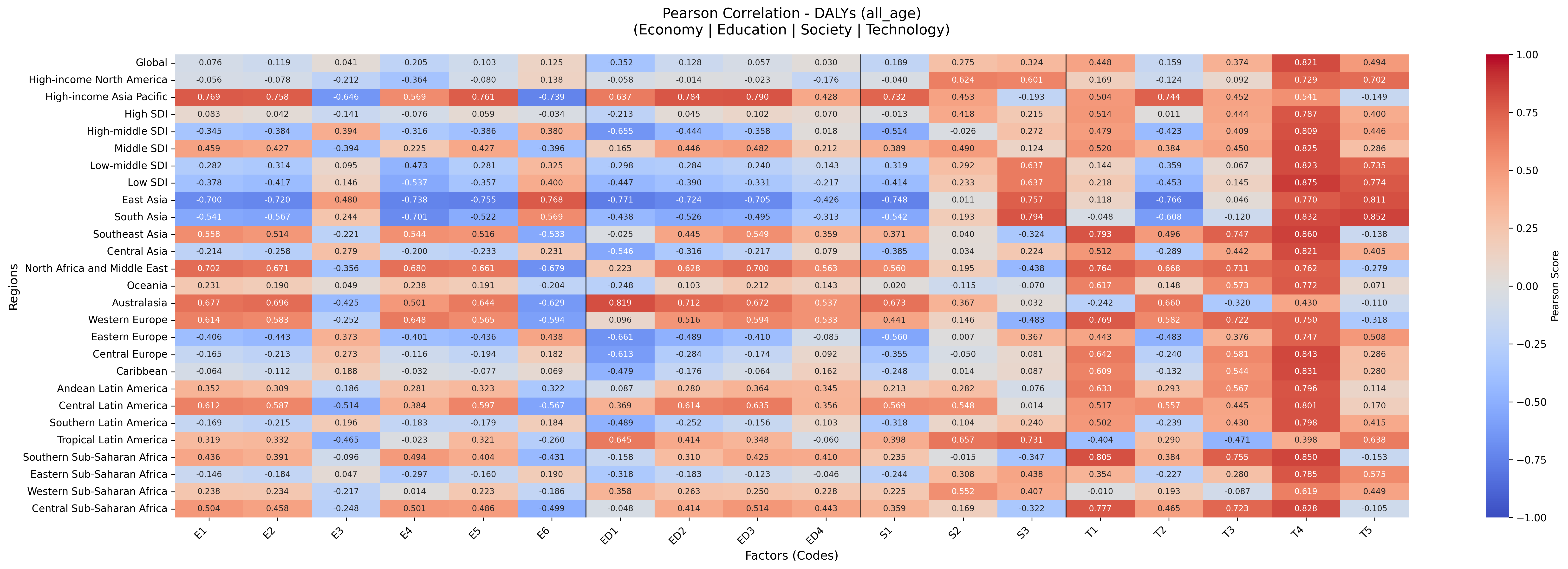}
\caption{Pearson Correlation Analysis - All Ages (DALYs). The heatmap shows correlation coefficients between socioeconomic factors (x-axis) and mental health burden across global regions (y-axis). Red indicates positive correlations, blue indicates negative correlations, and white indicates weak or no correlation.}
\label{fig:pearson_dalys_all}
\end{figure*}

The analysis of all age groups reveals complex regional patterns in the relationship between socioeconomic factors and mental health burden. Economic development indicators (E1-E6) show variable correlations across regions, with GDP per capita (E2) demonstrating strong negative correlations (r = -0.81) in East Asia, indicating that higher economic development is associated with lower mental health burden in this region. However, the same indicator shows positive correlations (r = 0.76) in Australasia, suggesting that the relationship between economic development and mental health is context-dependent and may follow different patterns in different regional contexts.

Educational factors (ED1-ED4) consistently demonstrate protective effects across most regions, with tertiary education enrollment (ED3) showing particularly strong negative correlations (r = -0.80) in East Asia and South Asia. This finding suggests that higher education levels are universally associated with better mental health outcomes, regardless of regional economic development levels. The consistent protective effect of education across diverse regional contexts highlights the importance of educational development as a universal strategy for improving mental health outcomes.

\begin{figure*}[htbp]
	\centering
\includegraphics[width=\textwidth]{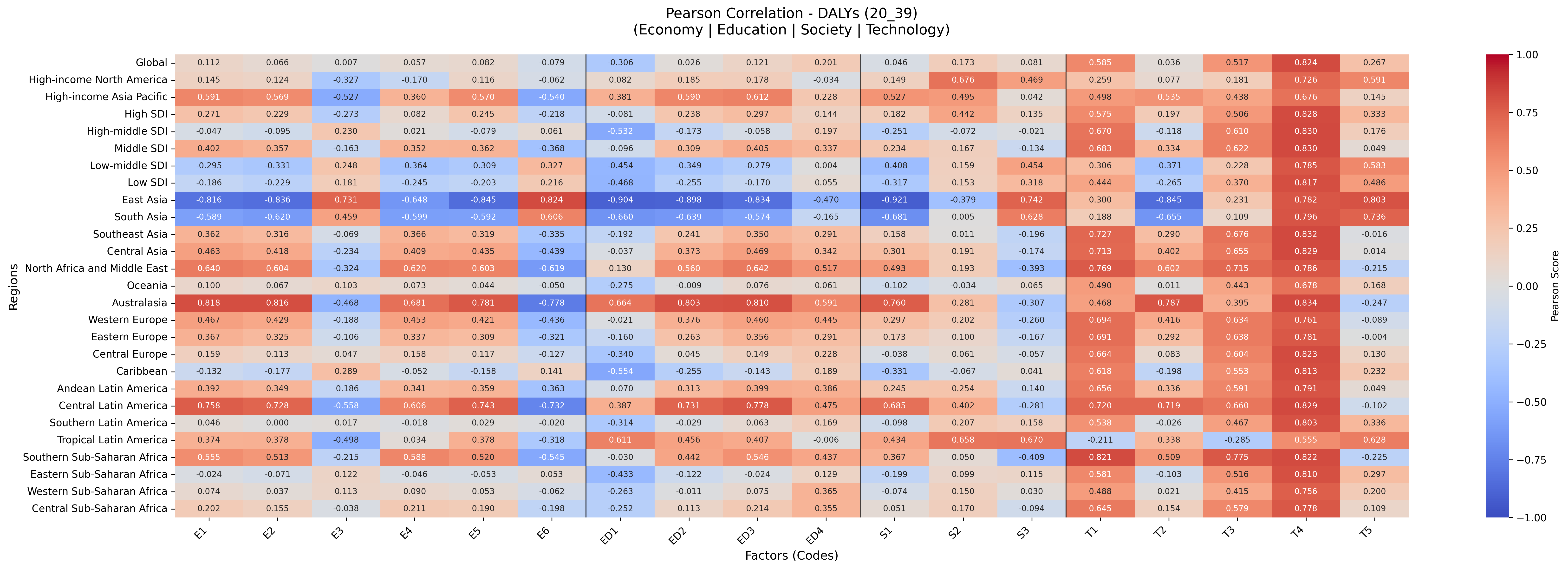}
\caption{Pearson Correlation Analysis - Young Adults (20-39 years). The heatmap shows stronger and more direct relationships between socioeconomic factors and mental health outcomes in the younger age group, with economic factors having immediate impacts and educational factors demonstrating the strongest protective effects.}
\label{fig:pearson_young_adults}
\end{figure*}

After analyzing young people aged between 20 and 39, it was found that compared with older people, there is a stronger and more direct connection between socio-economic factors and mental health outcomes. Economic factors have a direct impact, among which the unemployment rate, or S2, shows a strong positive correlation with mental health burdens. The correlation coefficient r is equal to 0.64. The impact of economic instability on the mental health of young people is particularly severe. Educational factors still demonstrate the strongest protective effect. All educational indicators, namely ED1 to ED4, show a consistent negative correlation with mental health burden. This finding indicates that educational opportunities and achievements are very crucial for the mental health protection of young people. It is possible to achieve this by enhancing skills in dealing with various problems, improving employment prospects, and strengthening social support networks.

\subsection{Mutual Information Analysis}

Mutual information analysis can capture both linear and nonlinear relationships between socio-economic factors and mental health outcomes, thereby providing insights into complex interactions that may not be easily discovered through linear correlation methods. This non-parametric analytical approach is crucial for identifying the threshold effects of the relationship between socio-economic development and mental health. And those complex multi-factor interactions are particularly valuable.

\begin{figure*}[htbp]
	\centering
\includegraphics[width=\textwidth]{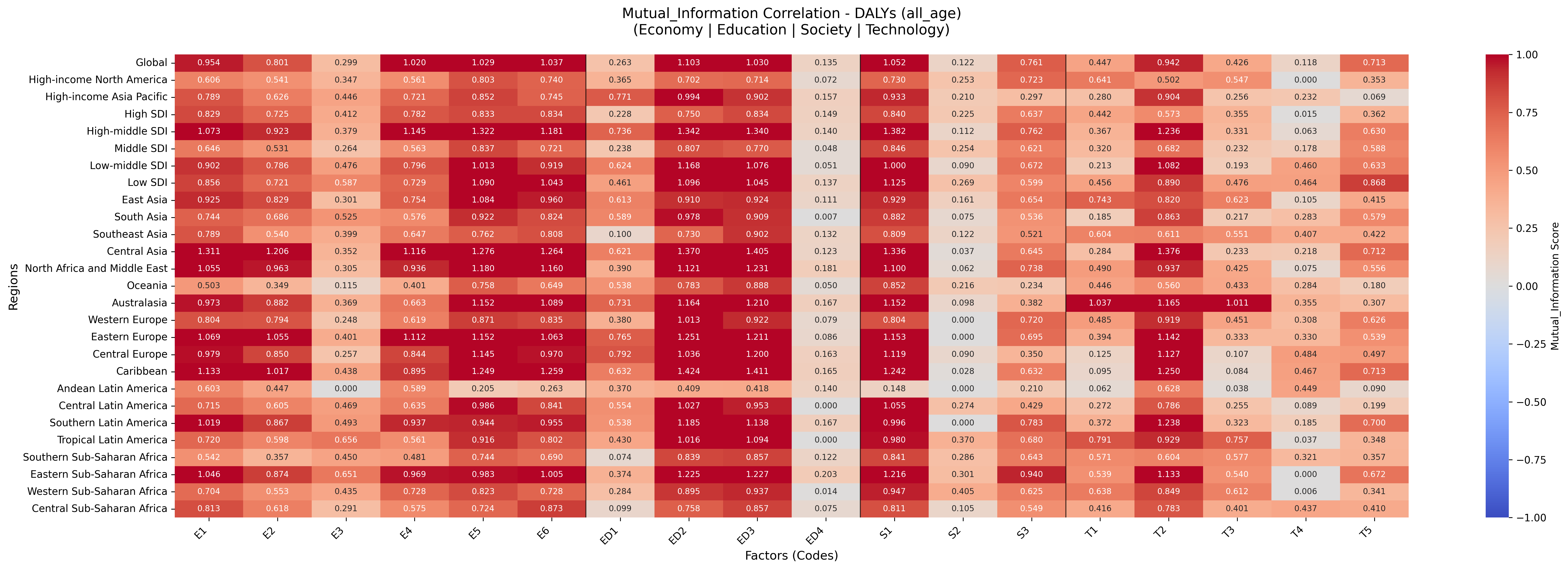}
\caption{Mutual Information Analysis - All Ages (DALYs). The heatmap shows mutual information values between socioeconomic factors (x-axis) and mental health burden across global regions (y-axis). Higher values indicate stronger associations, including non-linear and threshold relationships.}
\label{fig:mi_dalys_all}
\end{figure*}

Mutual information analysis has revealed the threshold effect of economic development and the complex nonlinear interaction among multiple factors. The relationship between per capita GDP and mental health outcomes can be seen from the analysis that the relationship between economic development and mental health is not a simple linear one, but follows a threshold pattern. In this model, the correlation between moderate economic development and mental health outcomes is the strongest. This finding implies that the benefits that economic development brings to mental health may no longer increase when the development level is very high, and may even have adverse effects instead. This might be because the pressure has increased and people have become socially isolated. It could be caused by factors such as changes in lifestyle related to highly developed economies.

There is a particularly complex relationship between technical factors (T1-T5) and mental health outcomes. Mobile cellular subscriptions (T2) show a strong association with mental health outcomes in certain regions, while fixed broadband users (T3) exhibit different patterns in the association with mental health outcomes. This complex situation indicates that The impact of technology adoption on mental health depends on the specific type of technology and the regional context in which it is located. Some technologies may provide social connections and support, but others may lead to social isolation or information overload.

\subsection{Granger Causality Analysis}

Granger causality analysis examines the temporal relationship between socio-economic factors and mental health outcomes. It can provide some insights into the direction and temporal aspects of the causal relationship. This analytical method is effective in understanding how changes in socio-economic conditions affect mental health outcomes over time and in determining the best timing for policy intervention. It has particularly great value.

\begin{figure*}[htbp]
	\centering
\includegraphics[width=\textwidth]{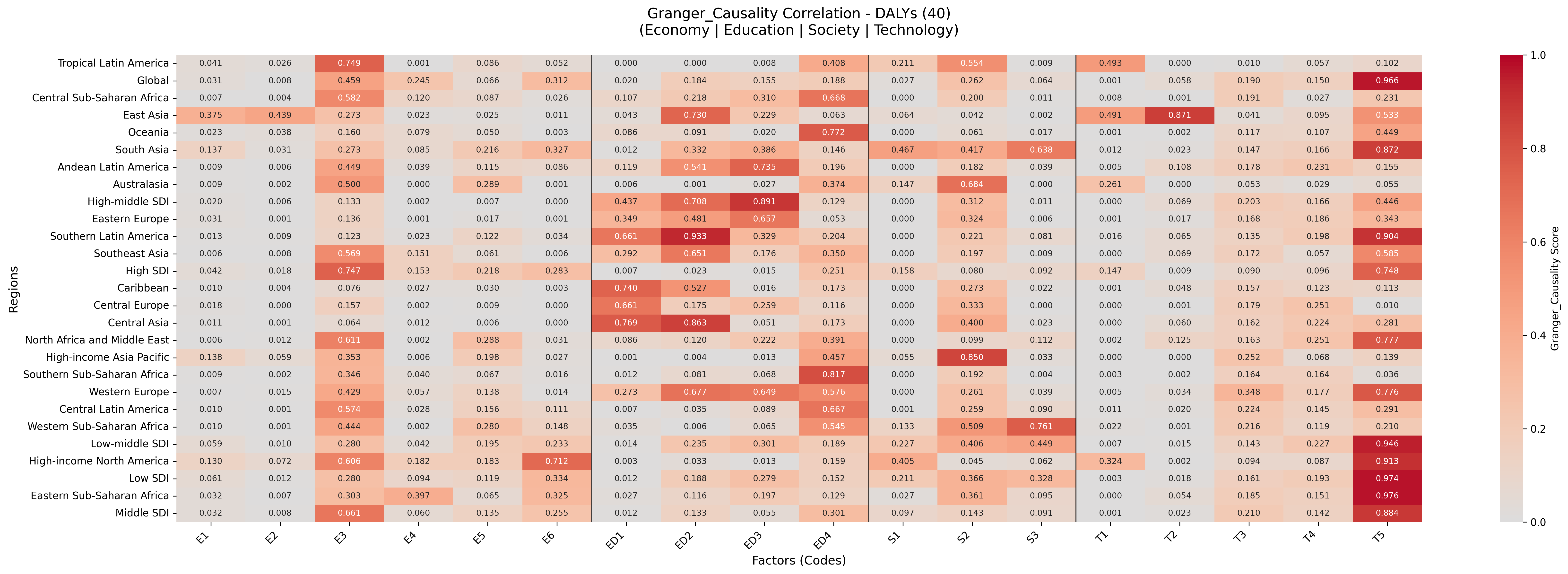}
\caption{Granger Causality Analysis - Older Adults (40+ years). The heatmap shows Granger causality test results between socioeconomic factors (x-axis) and mental health burden across global regions (y-axis). Lower p-values indicate stronger temporal relationships, with factors showing causal effects on mental health outcomes.}
\label{fig:granger_older_adults}
\end{figure*}

The Granger causality analysis reveals distinct temporal patterns in the relationship between socioeconomic factors and mental health outcomes, with different factors showing different lag structures. Among numerous socio-economic factors, for instance, changes in per capita GDP (E2) can cause a lag of 1 to 2 years in the mental health outcomes resulting from Granger causality. This indicates that economic changes have a relatively direct impact on mental health. From this discovery, it can be known that economic policies aimed at improving mental health outcomes may show certain effects within one to two years. In this case, these economic policies become particularly valuable for short-term intervention strategies.

The impact of educational factors on mental health outcomes takes a relatively long time. After changes in educational indicators (ED1-ED4), the mental health outcomes caused by Granger will lag by 3 to 5 years. This relatively long lag structure reflects that it takes some time for investment in education to be transformed into better mental health outcomes through certain mechanisms, including enhancing coping skills, improving employment prospects, and strengthening social support networks. The long-term nature of educational effects clearly indicates that If education is regarded as a long-term strategy for improving mental health outcomes, continuous investment in it is necessary.

The results of Granger causality analysis show that there is a rather prominent time pattern in the relationship between socio-economic factors and mental health outcomes. Different socio-economic factors exhibit different lag structures. Take the change in per capita GDP as an example; the mental health outcomes it causes lag by 1 to 2 years. Economic changes have a relatively direct impact on mental health. From this discovery, it can be inferred that economic policies aimed at improving mental health outcomes may show certain effects within one to two years. These economic policies thus become particularly valuable for short-term intervention strategies.

The impact of technology adoption on mental health outcomes can be seen quickly, and the lag time of changes in mental health outcomes caused by the technical indicator (T1-T5) Granger is the shortest. This direct impact indicates that technical policies may have a rapid influence on mental health outcomes through certain mechanisms, including improving access to mental health services, strengthening social connections, or increasing information and support resources, etc.

\subsection{Maximal Information Coefficient (MIC) Analysis}

MIC analysis has identified complex nonlinear relationships between socio-economic factors and mental health outcomes, as well as the interactions among multiple factors. This analytical approach is particularly valuable for revealing threshold effects and complex patterns that may not be detectable by other analytical methods.

\begin{figure*}[htbp]
	\centering
\includegraphics[width=\textwidth]{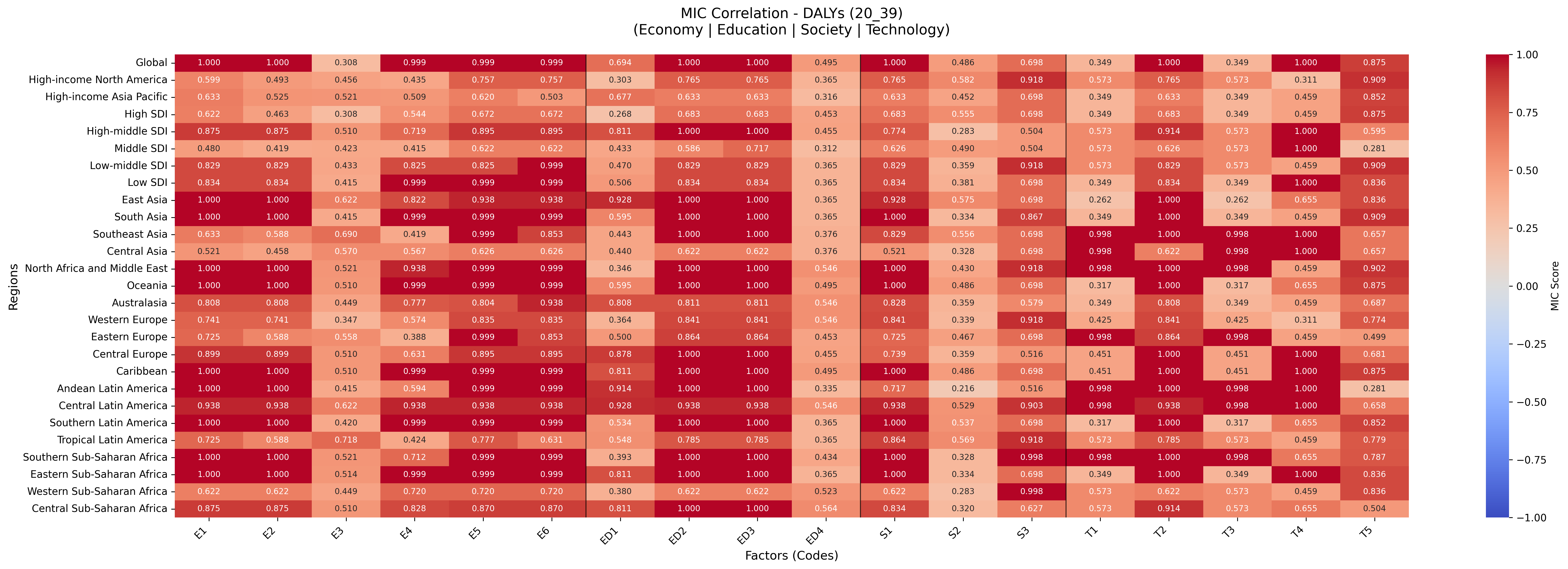}
\caption{MIC Analysis - Young Adults (20-39 years). The heatmap shows MIC values between socioeconomic factors (x-axis) and mental health burden across global regions (y-axis). Higher values indicate stronger complex associations, including non-linear and threshold relationships.}
\label{fig:mic_young_adults}
\end{figure*}

MIC analysis revealed a non-monotonic relationship, in which the association between moderate development and mental health outcomes was the strongest. This indicates that the relationship between socio-economic development and mental health is not a simple linear one but follows a more complex pattern. This finding implies that mental health outcomes are not determined by a single indicator. Rather, it depends on the combination of various factors. This also highlights that for mental health policies, it is crucial to adopt a comprehensive approach that simultaneously addresses multiple socio-economic factors.

The analysis of young adults (20-39 years) shows particularly complex patterns, with multiple factors demonstrating strong interactions. Economic factors (E1-E6) show complex relationships with mental health outcomes, with some factors demonstrating protective effects while others show risk effects depending on the specific combination of factors present. This complexity suggests that the relationship between socioeconomic factors and mental health in young adults is highly context-dependent and requires careful consideration of multiple factors when developing intervention strategies.

In the MIC analysis, educational factors, such as ED1-ED4, continue to demonstrate a very strong protective effect. However, the analysis results show that when educational factors are combined with other factors such as economic stability and social support, these effects will be enhanced. This finding indicates that The best results are achieved when educational intervention is combined with other socio-economic interventions, which also highlights the crucial importance of adopting a comprehensive approach for mental health policies.

\section{Conclusion}

This paper finds that socio-economic factors have a very prominent impact on the mental health status of different age groups and in various regions around the world. In all regions, educational factors have always played a protective role in mental health. There is a strong negative correlation between the enrollment rate of higher education and the burden of mental health. That is to say, the higher the enrollment rate of higher education, The lighter the mental health burden. The relationship between economic development and mental health has regional characteristics. For different regional conditions, methods suitable for local areas should be adopted. Compared with the elderly over 40 years old, young people aged between 20 and 39 have a stronger and more direct connection between their socio-economic factors and mental health status. There is a particularly strong positive correlation between the unemployment rate and the mental health burden. The higher the unemployment rate, the heavier the mental health burden. By analyzing time, it can be known that economic changes will have an impact on mental health within one to two years. However, for educational factors to play a role, a longer-term investment is needed, approximately three to five years. The application of technology has a direct impact on mental health. The research results show that Mental health status is not determined by a single indicator but is the result of the combined effect of multiple factors. This indicates that formulating comprehensive policies is extremely crucial.

This paper's analysis is limited by aggregate data that may not capture individual-level variations and missing data in some socioeconomic indicators, particularly in low-income regions. Future research should investigate causal mechanisms through longitudinal studies and examine individual-level data to understand how aggregate patterns translate to personal outcomes, while developing region-specific intervention strategies that address the complex interactions between socioeconomic factors and mental health outcomes.

\bibliographystyle{IEEEtran}
\bibliography{main}

\end{document}